\documentclass[10pt]{article}
\usepackage{graphicx}
\usepackage{amssymb}
\usepackage{atlasphysics}

\def\Title#1{\begin{center} {\Large #1 } \end{center}}
\def\Author#1{\begin{center}{ \sc #1} \end{center}}
\def\Address#1{\begin{center}{ \it #1} \end{center}}

\newcommand\pubblock{\rightline{\begin{tabular}{l} Proceedings of the Second Annual LHCP\\ \pubnumber\\
         \pubdate  \end{tabular}}}

\newenvironment{Abstract}{\begin{quotation} \begin{center} 
             \large ABSTRACT \end{center}\bigskip 
      \begin{center}\begin{large}}{\end{large}\end{center} \end{quotation}}

\newenvironment{Presented}{\begin{quotation} \begin{center} 
             PRESENTED AT\end{center}\bigskip 
      \begin{center}\begin{large}}{\end{large}\end{center} \end{quotation}}

\def\beq{\begin{equation}}
\def\eeq#1{\label{#1}\end{equation}}
\def\eeqn{\end{equation}}

\def\beqa{\begin{eqnarray}}
\def\eeqa#1{\label{#1}\end{eqnarray}}
\def\eeqan{\end{eqnarray}}

\def\sst{\scriptscriptstyle}

\let\bar=\overbar

\def\Dslash{\not{\hbox{\kern-4pt $D$}}}
\def\dslash{\not{\hbox{\kern-2pt $\del$}}}

\def\msb{{\bar{\ssstyle M \kern -1pt S}}}

\usepackage{color}
\usepackage{lineno}

\newcommand\pubnumber{ATL-PHYS-PROC-2014-124}

\newcommand\pubdate{\today}

\def\affiliation{
On behalf of the ATLAS Collaboration, \\
Max-Planck-Institut f\"ur Physik\\
F\"ohringer Ring 6, 80805 M\"unchen, Germany}

\def \met {\mbox{${\not}{E_T}$} }

\def\fb1{~fb$^{-1}$}

\def\gc{~GeV$/c$}
\def\gc2{~GeV$/c^{2}$}

\newcommand{\rpvlambda}[1][ijk]{\ensuremath{\lambda_{#1}}}

\begin{document}

\large
\begin{titlepage}
\pubblock

\vfill
\Title{Searches for electroweak production of supersymmetric gauginos and sleptons with the ATLAS detector}
\vfill

\Author{Federico Sforza\\ {\small }}
\Address{\affiliation}
\vfill
\begin{Abstract}
Many supersymmetry models feature gauginos and also sleptons with masses below a few hundred GeV. These can give rise to direct pair production rates at the LHC that can be observed in the data sample recorded by the ATLAS detector. The talk presents results from searches for gaugino and slepton pair production in final states with leptons.
\end{Abstract}
\vfill

\begin{Presented}
The Second Annual Conference\\
 on Large Hadron Collider Physics \\
Columbia University, New York, U.S.A \\ 
June 2-7, 2014
\end{Presented}
\vfill
\end{titlepage}
\def\thefootnote{\fnsymbol{footnote}}
\setcounter{footnote}{0}
%

\normalsize 


\section{Introduction}

The theory of Supersymmetry  (SUSY) is one of the favored extensions of the Standard Model (SM) and considerable effort has been devoted by the LHC experiments to the observation of the new predicted particles. Experimental constraints up to the TeV scale have been set if the SUSY particles are produced via the strong interaction (i.e. gluino and squark production). However, in the case that squark and gluino masses are out of experimental reach, the electroweak (EW) production of SUSY particles, which is expected to have lower production cross section at the LHC, may be the key for the discovery of SUSY. In this scenario, weaker experimental constraints exist.

Signatures with leptons are the favored analysis channel for the EW SUSY searches. This paper summarizes the results obtained in such channels by the ATLAS~\cite{atlasDet} collaboration on a 20\fb1 dataset of $pp$ collisions at $\sqrt{s}=8$~TeV.

\section{Phenomenological introduction and search strategy}

The EW SUSY sector comprises the SUSY partners of the SM leptons (sleptons $\slepton$),  neutrinos (sneutrinos $\tilde{\nu}$), EW gauge bosons (gauginos), and  Higgs bosons (higgsinos). In the framework of the  Minimal Supersymmetric Model (MSSM), the higgsinos and the gauginos form two charged and four neutral mass eigenstates, respectively named  $\chinoonepm, \chinotwopm$ (or $C1, C2$) and $\ninoone-\ninofour$ (or $N1 - N4$), in order of increasing mass. At a center of mass energy of 8 TeV, the predicted production cross section for charginos and neutralinos of $\approx 400$~GeV mass is of the same magnitude as the one for strong production of SUSY particles of $\gtrsim 1$~TeV mass. 

\begin{figure}[!htb]
\centering
\includegraphics[width=0.32\textwidth]{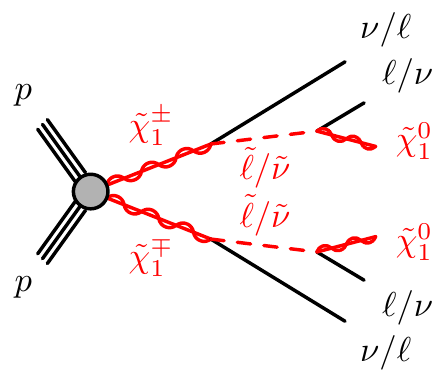}
\includegraphics[width=0.32\textwidth]{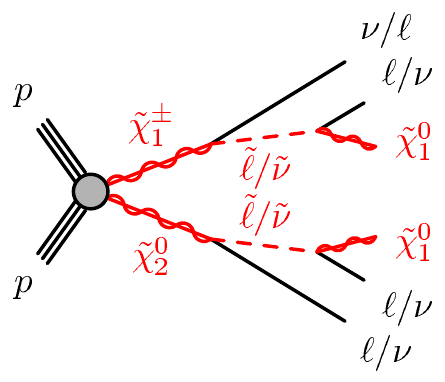}
\includegraphics[width=0.32\textwidth]{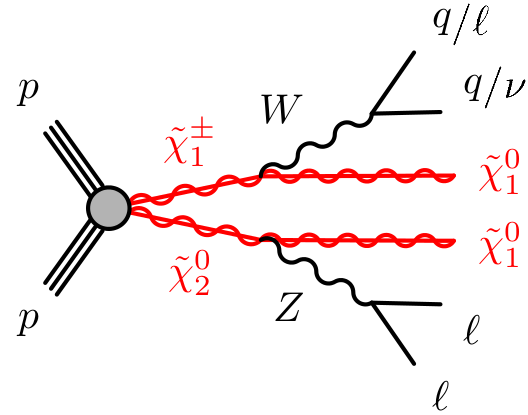}
\caption{Diagrams showing three examples of EW SUSY production and decay. On the left: C1C1 production with slepton mediated decays to a 2-lepton final state. On the center: C1N2 production with slepton mediated decays to a 3-lepton final state. On the right: C1N2 production with $WZ$ mediated decay to a 2-lepton (plus 2-jets) final state or to a 3-lepton final state.}
\label{fig:fey}
\end{figure}

Figure~\ref{fig:fey} shows three examples of production and decay of EW SUSY particles: these are  pair produced and then decay to lepton rich final states plus the lightest supersymmetric particle (LSP) which, in the case of R-parity conservation, is stable and it is detected as missing transverse energy ($\met$) in the collision. The analysis search strategies are based on the identification of clean leptonic final states and, depending on the lepton multiplicity, different new-physics scenarios can be examined.  Because of the vastness of the MSSM parameter phase space, the analysis optimization is usually done through simplified models targeting the direct production of specific particles (e.g. C1C1 or C1N2 in Figure~\ref{fig:fey}) and specific cascade decays where the only free parameters are the masses of the relevant particles and their decay modes (e.g. via $WZ$ or slepton in Figure~\ref{fig:fey}). The results are then reinterpreted in the phenomenologically constrained MSSM~\cite{pMSSM}.

\section{Search results based on lepton signatures}

The ATLAS collaboration has recently released five analyses targeting different final states: with one light  ($e$ or $\mu$) lepton~\cite{1lep}, two light leptons~\cite{2lep}, two hadronically decaying $\tau$~\cite{2tau}, three leptons~\cite{3lep}, and four leptons~\cite{4lep}.

The physics object identification criteria (i.e. leptons, jets, $b-$jets, $\met$, etc.) are similar between the analyses to ease the comparison between them and to allow the combination of the results. Usually signal enriched regions (SR) are selected by applying tight cuts on the kinematic variables and on the kind of reconstructed objects. For example the top background can be suppressed by applying a veto on $b-$tagged jets and/or a low amount of hadronic activity in the events, or $Z-$boson associated backgrounds can be suppressed vetoing events with reconstructed $Z-$candidates ($Z-$veto). 

The final background estimate is analysis dependent but it is usually fully or partially based on Monte-Carlo for the well defined SM process (like diboson or top production), and based on data-driven techniques for backgrounds due to non-prompt or misidentified leptons that may be hard to simulate correctly. The background estimate is validated in special regions (named validation regions) which usually have a topology close to the SR but enriched in SM background events and depleted in signal.

\subsection{Single lepton analysis}
 Final states with one light lepton, $\met$ and 2 $b-$jets are selected in this analysis~\cite{1lep}. For the first time in this analysis the constraint of $m_h= 125$~GeV has been used  to optimize the signal selection as it is possible to target EW SUSY scenarios mediated by the $Wh$ production if the two $b-$jets are supposed to come from the decay of a light Higgs boson. For a massless $\ninoone$ mass ranges of $125 < m_{\chinoonep,\ninotwo} < 141$~GeV and $166 < m_{\chinoonep,\ninotwo} < 287$~GeV are excluded at 95\% confidence level.

\subsection{Two lepton analyses}\label{sec:2lep}

\begin{figure}[!htb]
\centering
\includegraphics[width=0.56\textwidth]{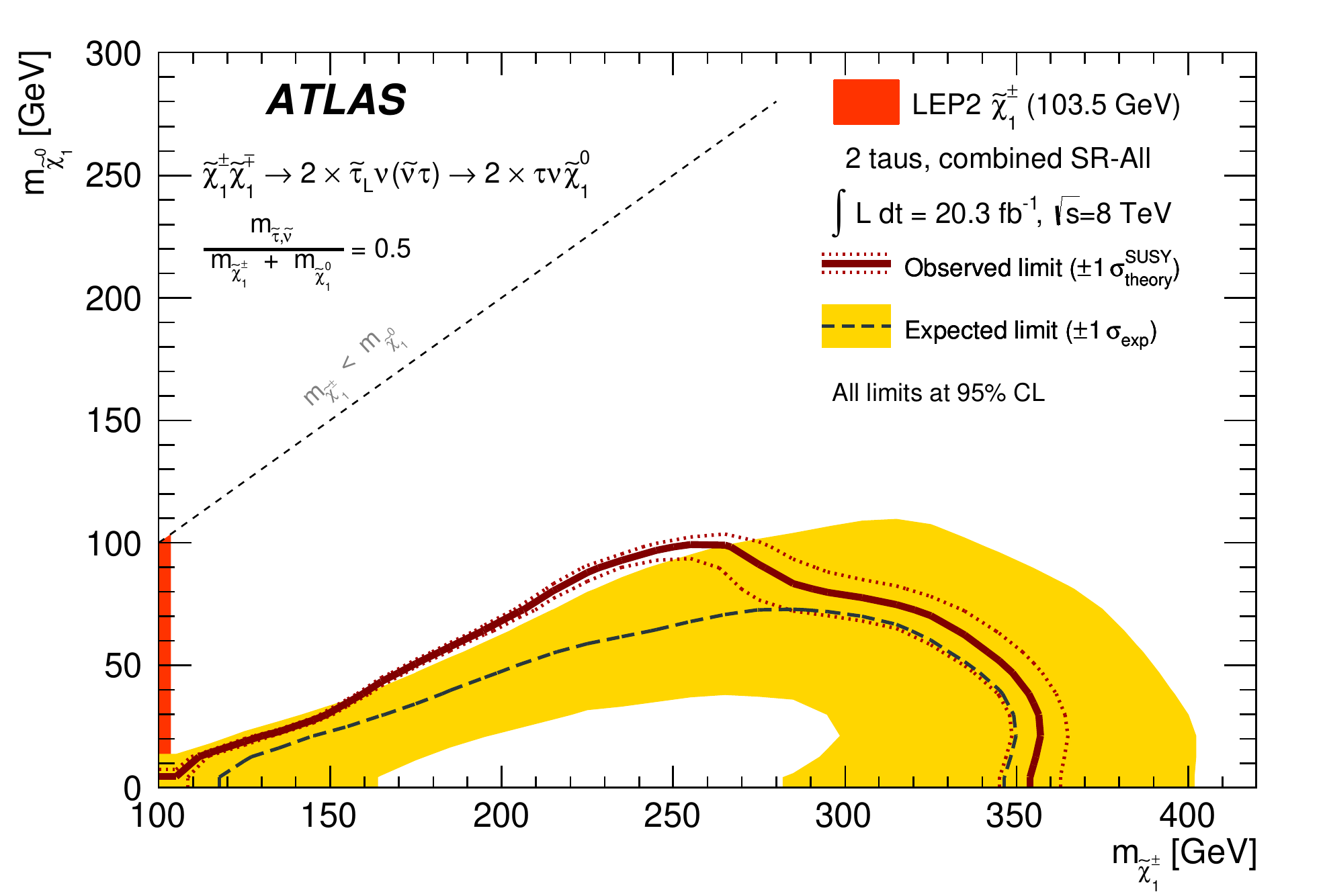}
\includegraphics[width=0.43\textwidth]{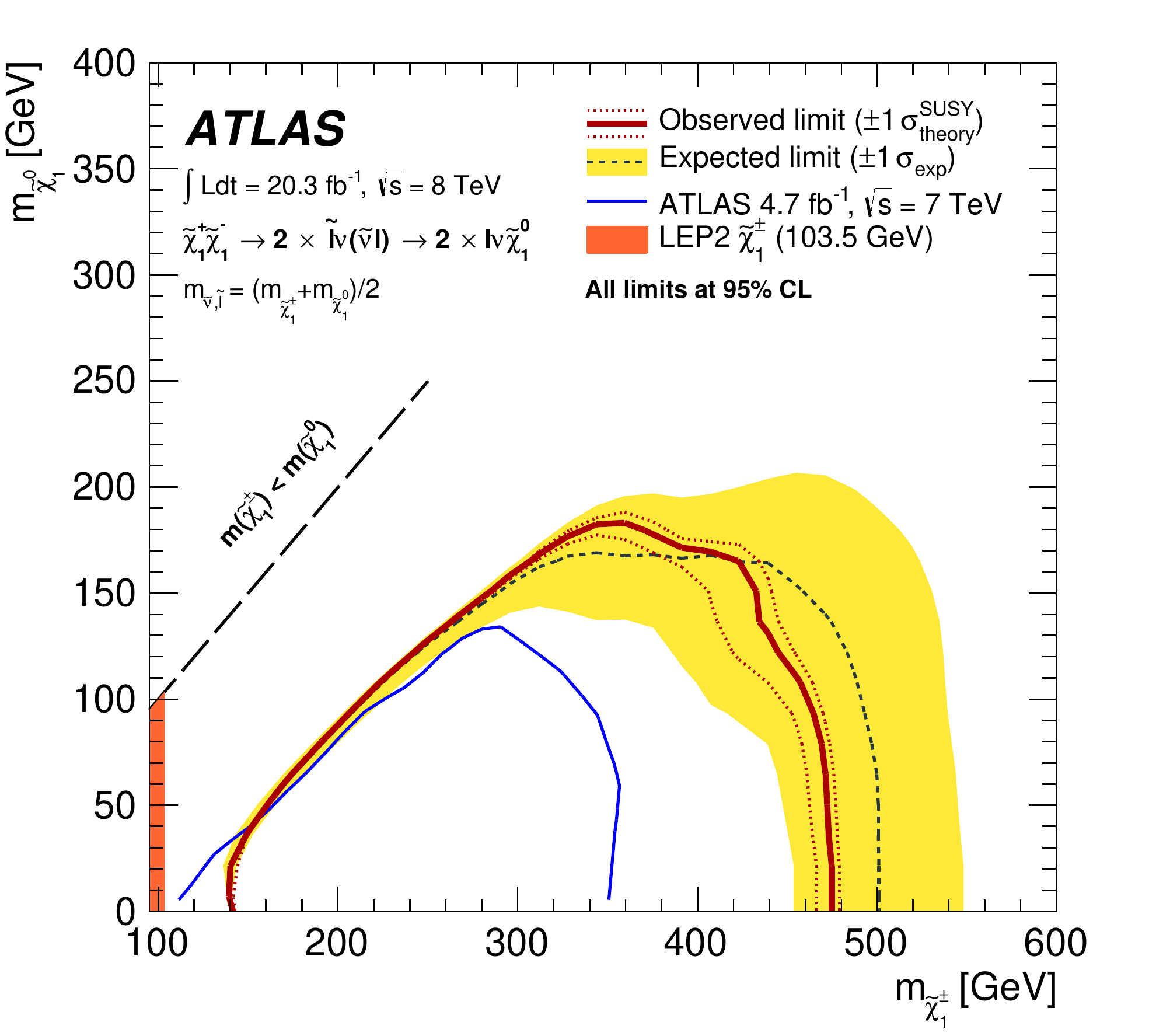}
\caption{Two examples of the 95\% confidence level mass exclusion limits for C1C1 production mediated by stau~\cite{2tau} (left) or by slepton~\cite{2lep} (right).}
\label{fig:slepMed_2lep_excl}
\end{figure}

Two different analyses target final states with, respectively, two light leptons~\cite{2lep} or two hadronically decaying $\tau$~\cite{2tau}. The first is mainly sensitive to scenarios with decays mediated by sleptons, $WW$, or $WZ$, the latter to scenarios with $\stau$ mediated decays. Both the analyses exploit the {\em stransverse  mass}~\cite{stransverse} ($m_{T2}$) variable to efficiently suppress SM backgrounds. Because of the variety of scenarios addressed in the two light lepton analysis, 7 specific SRs have been optimized using lepton flavor,  presence or absence of high-$\pt$ jets,  and different $m_{T2}$  thresholds.  
Figure~\ref{fig:slepMed_2lep_excl} shows two examples of the obtained exclusion limits for case of C1C1 production mediated by $\stau$ or $\slepton$. The mass exclusion limits are lower in the first case because of the experimental challenge of hadronic-tau reconstruction and the higher background expected in this final state. 

\subsection{Three lepton analyses}

\begin{figure}[!htb]
\centering
\includegraphics[width=0.495\textwidth]{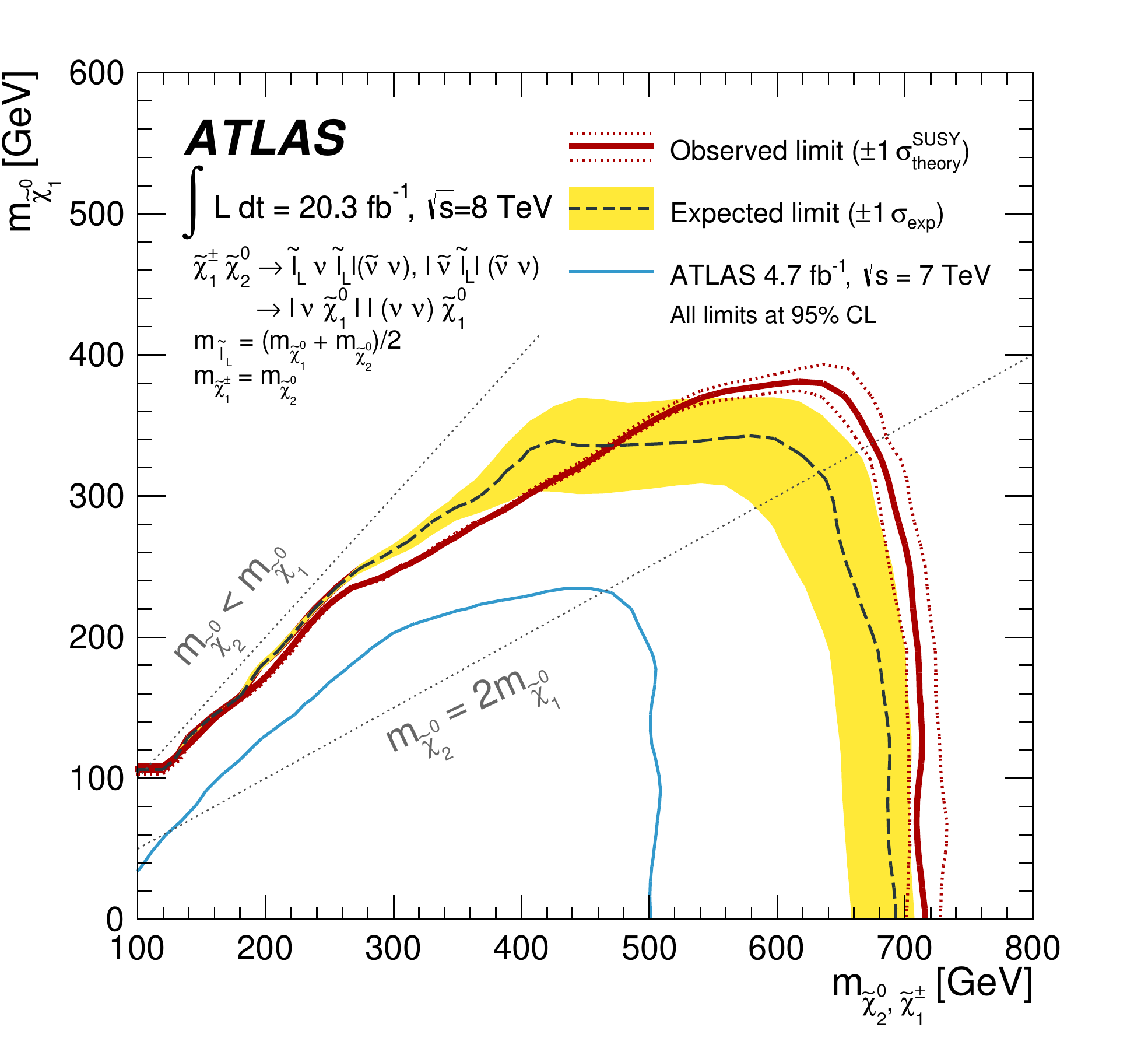}
\includegraphics[width=0.495\textwidth]{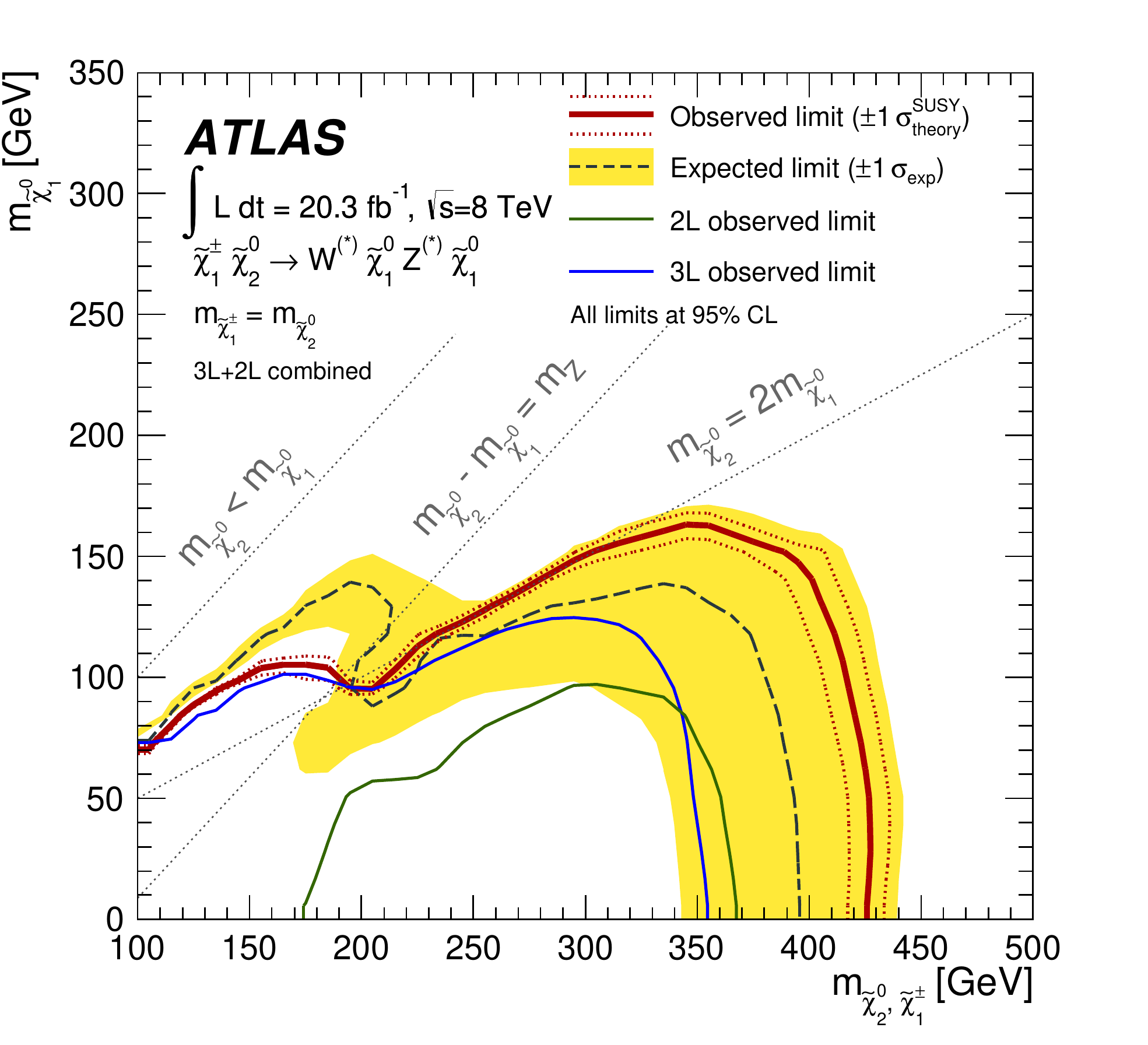}
\caption{Two examples of the 95\% confidence level mass exclusion limits for C1N2 production mediated by slepton~\cite{3lep} (left) or by $WZ$~\cite{2lep,3lep} (right). The second result is obtained from the combination of two different analyses targeting final states with two~\cite{2lep} or three~\cite{3lep} leptons in the final state: the green line indicates the central observed limit in the two lepton search alone, and the blue line indicates the central observed limit in the three-lepton search alone.}
\label{fig:wzMed_2_3lep_excl}
\end{figure}

The three lepton final state has the advantage of very low expected SM background and a rich topology where the multiplicity of physics objects can be used in advanced optimization of the SRs. This has been successfully done in the analysis described in~\cite{3lep} which is characterized by: one SR targeting $Wh$ mediated scenarios, 3 SRs with 1 or 2 hadronically decaying $\tau$ targeting $\stau$ mediated scenarios, and one binned SR where 20 independent bins are defined according to $\met$, mass of the same-flavor opposite-sign leptons and reconstructed transverse mass. Two examples of the obtained mass exclusion limits, depicted in Figure~\ref{fig:wzMed_2_3lep_excl}, show the strong sensitivity obtained for the $\slepton$ and $WZ$ mediated scenarios. Furthermore, the use of standardized physics object identification criteria and the possibility to investigate similar EW SUSY scenarios allows the sensitivity to $WZ$ mediated C1N2 production to be improved by statistically combining these results with the two-lepton analysis described in Section~\ref{sec:2lep} (the final result is also shown in Figure~\ref{fig:wzMed_2_3lep_excl}).

\subsection{Four lepton analysis}

The last analysis~\cite{4lep} described in this paper targets four-lepton final states (with up to two hadronically decaying $\tau$). As in the three-lepton analysis described in the previous section, the high multiplicity of physics objects and the low expected SM background are two points of strength of this search channel, however another relevant feature of this analysis is the possibility to target both R-parity conserving and R-parity violating (RPV) scenarios. In the RPV case, the LSP is unstable and allowed to decay to SM particles via the RPV super potential  $W_{\mathrm{RPV}}$. The part of it allowing leptonic final states reads:

\begin{equation}
  W_{\mathrm{RPV}}^{lep} = \frac{1}{2} \rpvlambda L_iL_j\bar{E_k}
\end{equation}
with $L_i$, $L_j$, and $E_k$ indicating the leptonic super-multiplets with both SM and SUSY fields and $i,j,k$ are generation indices related to the lepton flavor. Therefore  lepton rich final states are possible if $\rpvlambda \ne 0$. Several simplified models have been studied by assuming several choices of the Next-to-Lightest SUSY Particles (NLSP) and a $\ninoone$ LSP.

\begin{figure}[!htb]
\centering
\includegraphics[width=0.65\textwidth]{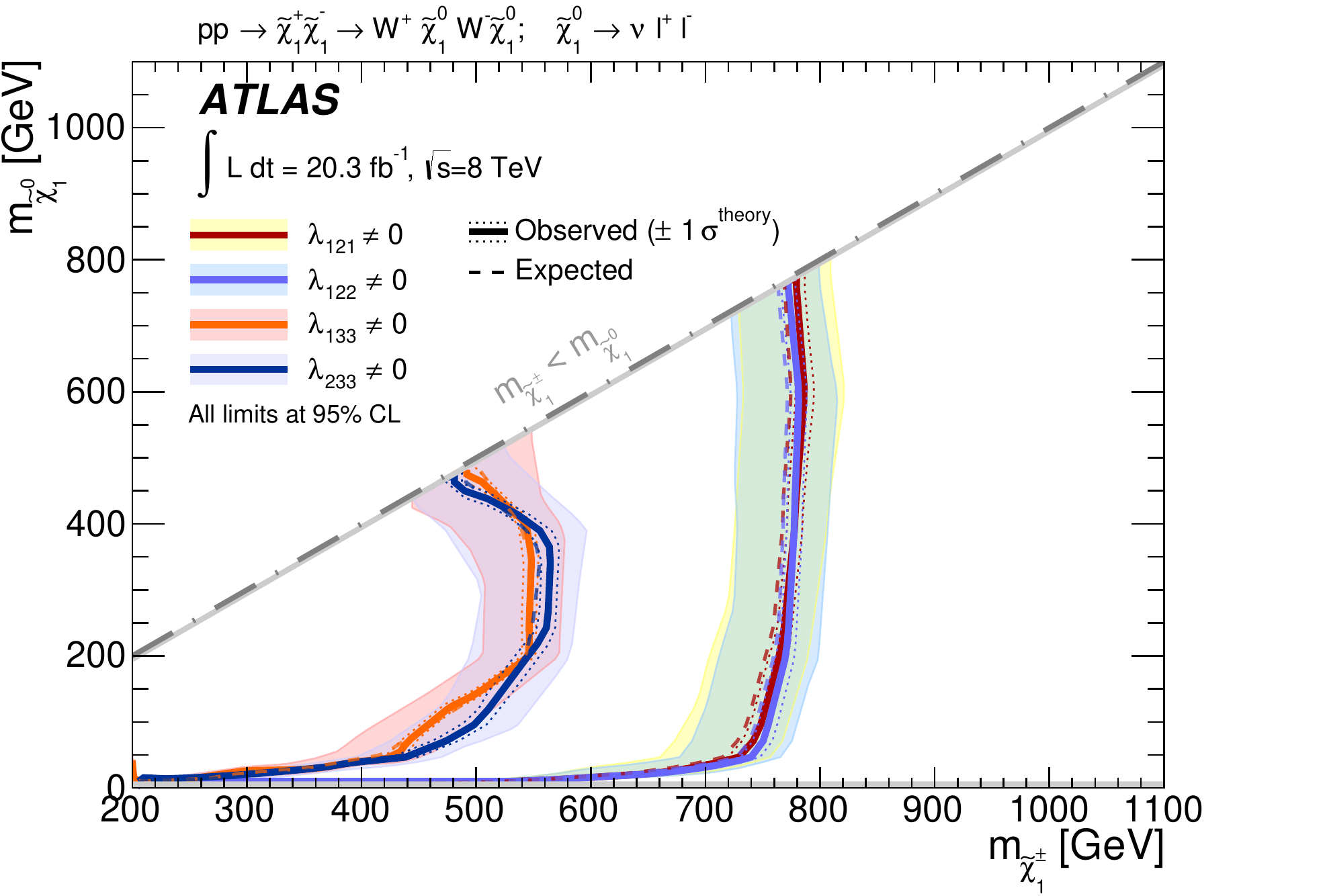}
\caption{Observed (solid) and expected (dashed) 95\%
CL exclusion limit contours for the RPV simplified models with a Wino-like chargino as the NLSP and assuming a promptly
decaying LSP. Each of the  $\rpvlambda$ corresponds to a specific lepton-flavor final state accessible to the LSP decay.}
\label{fig:rpv_expclusion}
\end{figure}

Nine SRs have been optimized targeting different production and decay modes. Figure~\ref{fig:rpv_expclusion} shows one example result: the 95\%
CL exclusion limits obtained  for RPV scenarios with a Wino-like chargino NLSP simplified models extend up to 750 GeV of mass. Limits on models with a gluino-like NLSP extend up to 1350 GeV of mass.

\section{Conclusions}
Electroweak production of SUSY particles is one possible channel for the discovery of SUSY at the LHC energy scale. The ATLAS collaboration has presented a variety of searches targeting the direct production of charginos, neutralinos and sleptons and exploiting the leptonic final states resulting from their decay. At present no evidence of new physics has been observed. 

\begin{figure}[!htb]
\centering
\includegraphics[width=0.65\textwidth]{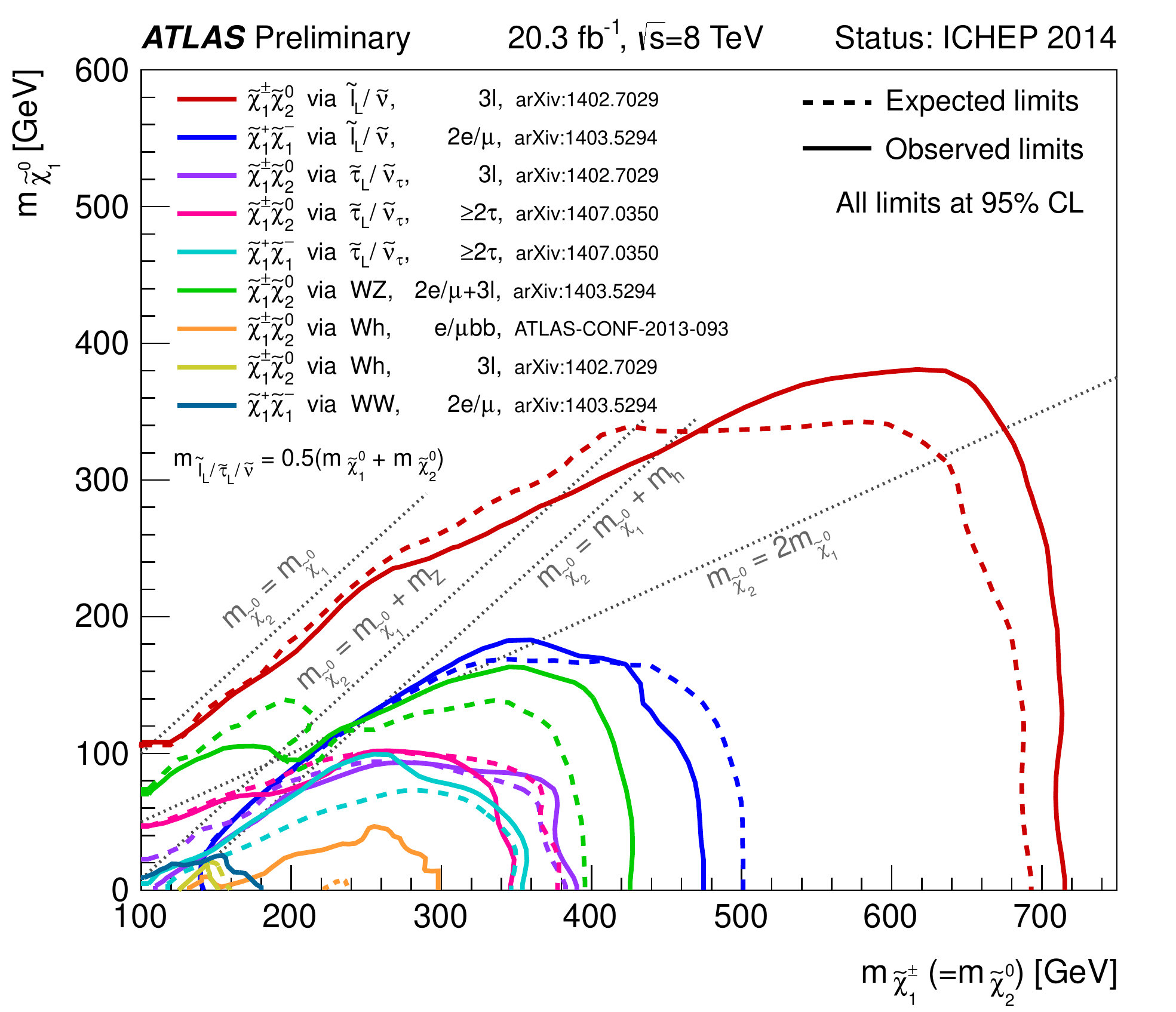}
\caption{Summary of ATLAS searches for electroweak production of charginos and neutralinos based on $20$~\fb1 of $pp$ collision data at $\sqrt{s} = 8$~TeV. Exclusion limits at 95\% confidence level are shown in the $\chinoonepm$, $\ninoone$ mass plane. The dashed and solid lines show the expected and observed limits, respectively, including all uncertainties except the theoretical signal cross section uncertainties. Several decay modes of the charginos and neutralinos are considered separately with 100\% branching fraction.}
\label{fig:ewk_summary}
\end{figure}

Figure~\ref{fig:ewk_summary} shows the summary of the current exclusion limits at 95\% CL~\cite{summary_plot} for charginos and neutralinos analyzed in a variety of simplified model scenarios. Using the $\chinoonepm$ and the $\ninoone$ as benchmark, 
for a $\chinoonepm$ mass of approximately $700$~GeV, $\ninoone$ masses up to $380$ GeV are excluded.
Simplified model scenarios with RPV decays of the LSP have been also investigated: the exclusion, in the case of decays to light leptons, is of the order of 750 GeV and 1350 GeV of mass for  charginos and gluinos respectively.

\end{document}